\documentclass[11pt]{article}
\usepackage[margin=3.5cm]{geometry}
\usepackage{ifpdf}
\usepackage{abstract} 
\usepackage{caption}
\usepackage{float}
\usepackage{multicol}
\usepackage{amsmath}
\usepackage{enumitem}
\usepackage{authblk}
\usepackage[normalem]{ulem}	

\usepackage{lineno}
\ifpdf 
    \usepackage[pdftex]{graphicx}   % to include graphics
    \usepackage[pdftex,     % sets up hyperref to use pdftex driver
            plainpages=false,   % allows page i and 1 to exist in the same document
            breaklinks=true,    % link texts can be{\LARGE {\tiny }} broken at the end of line
            colorlinks=true,
            citecolor=black,
            ]{hyperref} 
\else 
    \usepackage{graphicx}       % to include graphics
    \usepackage{hyperref}       % to simplify the use of \href
\fi 
\title{Understanding the connections between species distribution models}

\author[1]{Yan Wang \thanks{Correspondence author. Email:yan.wang@rmit.edu.au}} 
\author[1,2]{Lewi Stone}
\affil[1]{Discipline of Mathematical Sciences, School of Science, RMIT University, Melbourne, VIC, Australia}
\affil[2]{Biomathematics Unit, Department of Zoology, Faculty of Life Sciences, Tel Aviv University, Tel Aviv, Israel}

\usepackage{subcaption}
\usepackage{cleveref}

\usepackage{apacite}
\date{}
\begin{document}
\maketitle

\begin{abstract}
Models for accurately predicting species distributions have become essential tools for many ecological and conservation problems. For many species, presence-background (presence-only) data is the most commonly available type of spatial data.  A number of important methods have been proposed to model presence-background (PB) data, and there have been debates on the connection between these seemingly disparate methods. The paper begins by studying the close relationship between the  LI \cite{lancaster_case-control_1996} and LK \shortcite{lele_weighted_2006, royle_likelihood_2012} models, which were among the first developed methods for analysing PB data. The second part of the paper identifies close connections between the LK and point process models, as well as the equivalence between the Scaled Binomial (SB), Expectation-Maximization (EM), partial likelihood based Lele (2009) and LI methods, many of which have not been noted in the literature. We clarify that all these methods are the same in their ability to estimate the relative probability (or intensity) of presence from PB data; and the absolute probability of presence, when extra information of the species' prevalence is known. A new unified constrained LK (CLK) method is  also proposed as a generalisation of the better known existing approaches, with less theory involved and greater ease of implementation.

\end{abstract}


%\maketitle

\section{Introduction}
\label{s:intro}

Ecologists employ Species Distribution Models (SDMs)  to assist in  mapping the spatial distribution of  a species over its geographic range, despite  there being only limited observational data available.  These models often analyse presence-background (PB)  data (equivalently referred to as presence-only data), which contains  a  list  of `presences', or locations where individuals have been observed, but typically having  no information about absences - sites where species have not been observed. PB data is often plentifully available from so-called ``opportunistic surveys'' and can be found in museum and herbarium collections, historical database records \cite{pearce_modelling_2006}, and is now becoming increasingly available via online repositories such as the Global Biodiversity Information Facility (GBIF; http://www.gbif.org). Given such data, one of the key goals of SDMs is to estimate the site-specific probability of presence in the study region. Since SDMs assume that covariates are ultimately responsible for determining species' spatial distributions, SDMs model how covariates affect the local probability of presence.
To help estimate the site-specific presence probabilities, SDMs make use of extra background sites at which the information of the environmental covariates (e.g. temperature, altitude, etc.) are available.

More specifically, SDMs estimate the probability $p(y=1|x)$ that a species of interest is present, $y=1$ (versus absent, $y=0$) at a particular site, conditional on environmental covariates $x$ at that site.  The probability $p(y=1|x)$ is also referred to as  the resource selection probability function (RSPF). A common practice is to assume a parametric structure for modelling $p(y=1|x)$, for example, the widely used logit form, $\log\frac{p(y=1|x)}{1-p(y=1|x)}=\eta(x^T\beta)$. Here $\eta(x)$ can be a linear or a nonlinear function of $x$, and a logit-linear specification follows
\begin{equation}
\log\frac{p(y=1|x)}{1-p(y=1|x)}= \beta_0 + \sum_{i} \beta_i x_i .
\end{equation}
The goal of the SDM is to estimate all of the parameters $\beta_i$.

Also of interest is the species' overall prevalence $\pi$, which is the proportion of sites with species' presence in the study region. Thus, $\pi=\int p(y=1|x)dF(x)$,  
where $F(x)$ is the unknown probability distribution function for $x$. 
For PB data, $\pi$ is generally unknown since there is little or no information about the presence status of the background points.

A number of methods exist for modeling species distributions based on PB data, and in the paper we will focus on the maximum-likelihood based logistic regression methods discussed in Phillips and Elith (2013), the point process models (PPM) \shortcite{chakraborty_point_2011,warton_poisson_2010}, and the widely applied MAXENT method \shortcite{phillips_maximum_2006}. The logistic regression methods include the SC method \cite{steinberg_estimating_1992},  LI by \citeA{lancaster_case-control_1996}, LK by Lele and Keim (2006) and Royle et al. (2012), the Expectation-Maximisation (EM) of Ward et al. (2009), and the scaled binomial loss model (SB) by Phillips and Elith (2011). We also include the partial likelihood based Lele method by \citeA{lele_new_2009} in our study.

These methods on species distribution modelling have been developed independently using different definitions and framework. The key goal of the paper is to show the equivalence between these seemingly disparate models.  This is one of the major contributions of our manuscript. 
The connections are revealed initially by studying the close link between the LI and LK methods, which were among the first developed methods for analysing PB data. It will be shown for the first time that the LK method  is a numerical approximation of the LI method. 

Secondly, we examine the analogy between the PPM and the LK model, when the likelihood function of the PPM is approximated by its discrete counterpart. We also show the equivalence between the SB, EM, LI and the Lele methods. These equivalences have not been noted previously in the literature. 
Along with other findings on relations in the field, such as those done by \citeA{baddeley_2010,warton_poisson_2010,Aarts_2012,fithian_finite-sample_2013,renner_equivalence_2013}, we conclude that all these methods are essentially equivalent in their ability to estimate the relative probability of presence. Furthermore, we present a unified constrained LK (CLK) method, which bridges the gaps between these seemingly different approaches. Each of the methods discussed in the paper is shown to be a special case of the unified CLK method. 
\section{The relationship between LI and LK methods}
\label{s:LI/LK}

Lancaster and Imbens (1996) proposed a contaminated case control study for representing PB data, in which the set of sites in the study area is divided into two subsets. 
Subset 1 consists of all those sites in the study area on which the species is present. Subset 0 comprises the whole set of sites in the study area, with no information made available regarding which of these `background sites' the species is present or not. However, the relevant environmental covariates are known at all background sites.

LI defined a sequence of $n$ Bernoulli trials with the probability $h$ to choose between the presence (case) and background (contaminated controls) points. A binary indicator $u$ was used to denote the stratum, with $u=1$ if the observation was drawn from the presence, and $u=0$ if it was drawn from the whole population. After the $n$  Bernoulli trials, there are $n_1$ sites chosen with species presences, and $n_0$ background sites with unknown status.  The background points in analysing the PB data is usually taken as either a uniform sample or a regular grid with a sufficient large set of values of covariates. The distribution of the environmental covariates $F(x)$ can be approximated by a discrete distribution with unknown probabilities $\alpha_l$ on $L+1$ known points of support $x_l$ (Lancaster and Imbens, 1996). An empirical estimator of $\alpha_l$ is the fraction of observations taking the value $x_l$ in the background data, i.e., $\hat\alpha_l=n_l/n_0$. 

From Bayes theorem, we can derive $p(x|y=1)=\frac{p(y=1|x)f(x)}{\pi}$.
The joint distribution of stratum $u$ and covariates $x$ is: 
$g(x,u)=[p(x|y=1)h]^u[f(x)(1-h)]^{1-u}$ (Lancaster and Imbens, 1996) , which can be rewritten as $[\frac{p(y=1|x)f(x)h}{\pi}]^u[f(x)(1-h)]^{1-u}$ . The full likelihood function for the contaminated sampling scheme based on the joint distribution of $(x,u)$ is
\begin{eqnarray} 
L(\beta,h,\alpha,\pi) &= &\prod_{i=1}^{n}{[\frac{p(y_i=1|x_i,\beta)f(x_i)h}{\pi}]^{u_i} [f(x_i)(1-h)]^{1-u_i}} \nonumber \\
&= & \prod_{i=1}^{n} [\frac{p(y_i=1|x_i,\beta)}{\pi}]^{u_i} \prod_{i=1}^{n} f(x_i) \prod_{i=1}^n[h^{u_i} (1-h)^{1-u_i}] \nonumber\\
&=& L_1(\beta,\pi)*L_2(\alpha)*L_3(h), 
\label{eqn:LIfull}        
\end{eqnarray}
where the total number of sample points is $n=n_0+n_1$. 

By splitting the full likelihood into three partial likelihoods in (\ref{eqn:LIfull}), the role of each likelihood becomes clear. The partial likelihood function $L_3 (h)$, which is independent of other parts of the likelihood function, is used to estimate the unknown sampling proportion with a binomial type of estimator $\hat h=n_1/n$. Similarly, $L_2$ is relevant to the estimation of the probability distribution function of covariates $F(x)$. It is the partial likelihood $L_1(\beta,\pi$) that contributes to the estimation of $\beta$, and the probability of presence  $p(y=1|x,\beta)$. 

Let's take a further look at the partial likelihood of $L_1$. The population prevalence $\pi$ involves an integral $\int p(y=1|x)dF(x)$ , which can be approximated by $\sum_{x_l} p(y=1|x_l,\beta) \frac{n_l}{n_0}$ on $L+1$ known points of support $x_l$, with $F(x)$ replaced by its empirical estimate of $\hat\alpha_l=n_l/n_0$. This approximation for $\pi$ can be rewritten as $\frac{1}{n_0}\sum_{i=1}^{n_0} p(y=1|x_i,\beta)$, when we shift the sample space from the environmental space $x(s)$ to the geographic feature $s$ \cite{hastie_inference_2013}. Upon this transformation for $\pi$, the partial likelihood $L_1(\beta,\pi)$ in the LI method becomes
\begin{eqnarray}
L_1(\beta)=\prod_{i=1}^{n_1} \frac{p(y=1|x_i,\beta)}{\frac{1}{n_0}\sum_{j=1}^{n_0} p(y=1|x_j,\beta)}. 
\label{eqn:LK}
\end{eqnarray}
This approximation of the partial likelihood for $\beta$ is exactly the likelihood of the popular LK method \cite{lele_weighted_2006,royle_likelihood_2012}. 
Through maximising the likelihood, it is possible to estimate the best fitting parameters $\beta$ required to determine the probability of presence. The LK method can be viewed as a numerical approximation of the LI method, where the accuracy of the approximation will improve, in a statistical sense, by increasing the size of the background samples. In the following Simulations section, we will show the similarity between the LI and LK estimates, when the number of background points is large. 

Can the true probability of presence $p(y=1|x)$ be estimated from the LI or LK methods?  There have been extensive discussions on this topic \shortcite{lele_weighted_2006,ward_presence-only_2009,royle_likelihood_2012,phillips_estimating_2013,hastie_inference_2013,Solymos_Lele_2016}. 
In the Simulation section, we will demonstrate the need for extra information, such as the  parametric structure of $p(y|x)$ or prior knowledge of species' prevalence, $\pi$, in order to estimate the absolute probability of presence. 
As both the LI and LK methods require no prior knowledge of $\pi$, their successful operation relies on the resource selection probability function (RSPF) conditions, which have been given by \citeA{lele_weighted_2006} and further discussed in \citeA{Solymos_Lele_2016}. Loosely speaking, the RSPF condition includes, for example, that the true (actual) function of $\log p(y=1|x)$ is nonlinear, and not all covariates in the model are categorical.

\section{The relationship between LK and PPM}
\label{s:LK&maxent}

The point process model (PPM) in spatial analysis has recently been proposed as a versatile approach for analysing species presence-background  data \cite{warton_poisson_2010,chakraborty_point_2011}, because it treats space as continuous, which seems more realistic than discrete space approaches. Poisson point process models, however, have been shown to be closely connected to other popular methods in ecology such as MAXENT \cite{Aarts_2012,fithian_finite-sample_2013,renner_equivalence_2013}, logistic regression \cite{baddeley_2010,warton_poisson_2010} and resource selection models \cite{Aarts_2012}. \citeA{hefley_bias_2017} also showed that the discrete space SDMs can be linked to the PPM using the so-called change of support.

In this section, we will briefly demonstrate how the likelihood of the LK method is associated with the conditional likelihood of the PPM, which is equivalent to MAXENT. We note that \citeA{Aarts_2012} also observed the equivalence between the LK and the conditional PPM. However they did not provide any formal details of how the equivalence can be reached through a numerical approximation of the PPM, as we show here. 

In the PPM framework, PB data consists of a set of locations ${s_1,s_2,\cdots,s_{n_1}}$, where individuals of a species are observed in a region $D$. These locations are defined as a realization of a point process that is characterized by the intensity $\lambda(s)$, which varies spatially according to a parametric function of environmental features $x(s)$. The likelihood proposed for fitting an inhomogeneous Poisson process is
\begin{equation}
L(s_1,\cdots,s_{n_1},n_1)=\exp(-\int_D \lambda(s)ds)(\prod_{i=1}^{n_1} \lambda(s_i))/{n_1}! 
\label{eqn:full}
\end{equation}
\cite{cressie_statistics_2011,renner_point_2015}. This likelihood function was derived as the product of the conditional likelihood, 
\begin{equation}
L_c(s_1,\cdots,s_{n_1}|n_1)=\prod_{i=1}^{n_1}\frac{\lambda(s_i)}{\int_D \lambda(s)ds}, 
\label{eqn:con}
\end{equation}
and the marginal likelihood,
\begin{equation}
P(N(D)=n_1)=\exp(-\Lambda(D)) \Lambda(D)^{n_1}/n_1!
\end{equation}
\cite{Moller_Wagge_2004,dorazio_accounting_2014}, where $\Lambda(D)=\int_D \lambda(s)ds$ is the cumulative intensity over the study area of $D$. 

The likelihood (\ref{eqn:con}) involves an integral over a study area that cannot be computed exactly and must be approximated numerically. Berman and Turner (1992) developed a numerical quadrature method for estimating the integral by approximating it as a finite sum using any quadrature rule, i.e., $\int_D \lambda(s)ds \approx \sum_{i=1}^{n_0} \lambda(s_i) w_i$. In the simplest form, we assign equal weight to each quadrature point, for example $w_i=\frac{|D|}{n_0}$, by partitioning $D$ into $n_0$ equal rectangular tiles and a single quadrature point selected from each tile \cite{baddeley_spatial_2015}. $|D|$ represents the total area of the region. With this simple quadrature scheme, the conditional likelihood for the PPM is approximated by 
\begin{equation}
L_c=\prod_{i=1}^{n_1}\frac{\lambda(s_i)}{\frac{|D|}{n_0} \sum_{j=1}^{n_0} \lambda(s_{j})} \label{eqn:approx_con}.
\end{equation}

The above discretized version of the conditional likelihood of the  PPM is the same as the likelihood of MAXENT using the log-linear intensity function, $\log\lambda(s)=\beta'x(s)$ \cite{fithian_finite-sample_2013,renner_equivalence_2013}. We want to address that this approximation is also the analogy of the likelihood of the LK method in Eq.~(\ref{eqn:LK}). It is worth noting that the background points used to approximate $\pi$ in the LK method play exactly the same role as the quadrature points, which are used in the PPM for numerically evaluating the cumulative intensity $\int_D \lambda(s)ds$.  The choice of different quadrature schemes in approximating the conditional PPM can lead to models very different from the LK method (A discussion of the various quadrature schemes can be found in Chapter 9 of \citeA{baddeley_spatial_2015}).

The difference between the LK and the approximated version of the conditional PPM lies in the link functions being used for modelling $p(y|x,\beta)$ and  $\lambda(s)$ respectively, often chosen by consideration of the range of values a probability and an intensity function can take. Nevertheless, we will show through numerical simulations that the choice between the different link functions, for example, the logit, log-linear or the complementary loglog functions, makes little difference in estimating the ratio, $\frac{p(y=1|x_i,\beta)}{\pi}$, i.e., the relative probability (or intensity) of presence. In other words,  when using either the LK/LI, MAXENT or conditional PPM model in studying the PB data, all of them will yield the same relative probability (or intensity) of presence. It is also worth mentioning that although the PPM has been introduced as a natural framework for modelling PB data, its ability to produce the {\it relative} probability of presence (or relative intensity), which is free of grid or transect selection, is the same as other so-called discrete space models. 

Can the true intensity of presence be estimated with PPM methods? From the discussions of \citeA{fithian_finite-sample_2013}, \citeA{renner_point_2015} and \citeA{dorazio_predicting_2012}, we know that the PPM can only estimate the intensity of reported presence from its full likelihood function, instead of the intensity of true presence $\lambda(s)$.
It is because an underlying equation $\Lambda(D)=n_1$ is derived from the full likelihood function, in addition to the conditional likelihood of the PPM \cite{fithian_finite-sample_2013}. 
This additional information of $\Lambda(D)$ is biased, as the cumulative intensity should equal the number of true presence over the study area, whereas $n_1$ was only observed by opportunity.  %as it represents the prevalence of `species reporting' rather than 'species presence'. 
One way to correct for this bias is to make use of an appropriate species presence number in the PPM for $\Lambda(D)$.

\section{The relationship between EM, SB, Lele and LI methods}

\citeA{ward_presence-only_2009} proved that the probability of presence is not identifiable from PB data, if there was no information about the structure of the the probability function. Under this circumstance, the knowledge of the population's prevalence is required to estimate the true probability of presence. They used the commonly used logit function to fit the PB data, and proposed the Expectation-Maximisation (EM) algorithm to estimate the parameters of the logistic regression. The EM algorithm was able to estimate the probability of presence accurately at any site, using the species' prevalence as an additional information.  

Two other successful methods  discussed in the literature, the SC \cite{steinberg_estimating_1992} and SB \cite{Phillips_SB_2011} method, also require the true species' prevalence to obtain estimates of the site-specific probability of presence.  Although the EM and the SB methods work on different likelihood functions, their estimates of the probability of presence are essentially the same. It is also the first time to show the equivalence between the likelihood function of the EM/SB method and the LI method (see details in Appendix A). 

\citeA{lele_new_2009} proposed a new method, referred to as the Lele method in our paper, to improve the instability of the the LK method that they identified. The Lele method is a combination of the partial likelihood and data cloning to obtain the maximum likelihood estimator of both $\beta$ and $\pi$. In Appendix B, we show that the likelihood function of the Lele method is the same as that of the LI method, although  these two seemingly different approaches were developed independently. 

\section{General connections between all methods}

The methods discussed in this paper, i.e., LI, LK, Lele, EM, SC, SB, MAXENT, and the PPM, can be divided into three different camps, based on their underlying likelihood functions and type of extra information required. The LI, LK, and Lele methods are sorted into Camp 1, which can estimate the absolute probability of presence, provided that the probability function satisfies the RSPF conditions \cite{lele_weighted_2006,Solymos_Lele_2016}. 
The method of  EM, SB and SC  fall into Camp 2, and are also able to estimate the absolute site-specific probability of presence, using an additional input of the species' prevalence. The MAXENT, continuous space PPM method and its associated models, are categorized into Camp 3, according to their connection between one another \cite{warton_poisson_2010,baddeley_2010,Aarts_2012,fithian_finite-sample_2013,renner_equivalence_2013}.

\begin{figure}[htbp]
 \centerline{\includegraphics[width=6.5in]{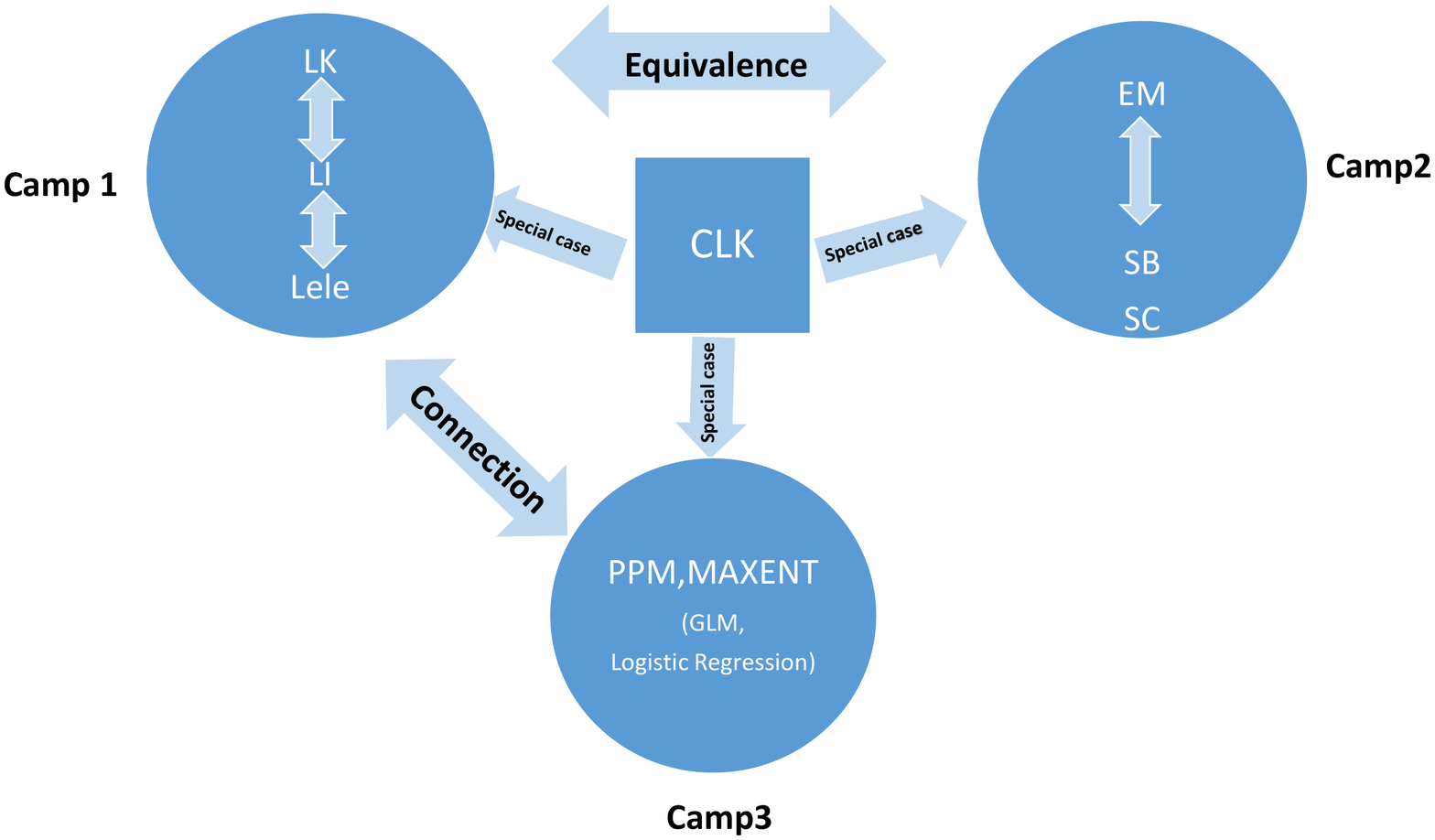}}
 \renewcommand{\baselinestretch}{1} \caption{The methods divide into three camps. Camp 1 includes the LI, LK and Lele methods that can estimate the probability of presence, given the RSPF conditions are satisfied. Camp 2 includes the EM, SC and SB methods that require the extra information of the species' population prevalence, in order to estimate the probability of presence. The MAXENT, PPM methods and its associates are included in Camp 3, which in general estimate the relative probability of presence or the probability of reported presence.}
\label{figure 1}
\end{figure}

We have discussed some of the pairwise relationships, such as between LI and LK, LK and PPM, SB and LI, and between Lele and LI, respectively. Is there a way to connect together all the methods discussed in the paper? 
We believe this is in fact possible and a summary of our findings is given in Fig.~\ref{figure 1}, where the relations inside the same camps and across different camps are first-time presented. 

%\newpage

The methods in Camp 1 and Camp 3 (e.g., the LK and the PPM) are shown to share a common conditional likelihood, which has the same structure as the partial likelihood $L_1(\beta,\pi)$ in Eq.~\ref{eqn:LIfull}. 
The methods in Camp 1 and 2 (e.g., the LI, Lele, EM and SB methods) are constructed on the same likelihood function, i.e., $L(\beta,h,\alpha,\pi)$ in Eq.~\ref{eqn:LIfull}, which can be further decomposed as the product of the likelihood $L_1(\beta,\pi)$ and other terms that do not involve both $\beta$ and $\pi$. Therefore, all the methods in the three camps are actually built on the same partial/conditional likelihoods, i.e.,  $L_1(\beta,\pi)$. In other words, all these seemingly different  SDM models are equivalent in their ability to estimate the {\it relative} probability of presence for modelling PB data, regardless of their different presentations.

The difference between Camp 1 and Camp 2 is that the methods in the latter require a pre-determined value of species' prevalence, $\pi$, while the LI and Lele methods in Camp 1 treats $\pi$ as an unknown parameter. As for the LK method, in order for the LI and Lele methods to identify $\pi$, the RSPF conditions listed in \citeA{lele_weighted_2006} and \citeA{Solymos_Lele_2016} need to be satisfied. However, this has led to controversy criticized by data scientists in particular \cite{ward_presence-only_2009,Phillips_SB_2011,hastie_inference_2013}, because the true parametric functions are generally unknown in practice, and the functions used to fit these true functions can be of different structures. Under these circumstances, a revised version of the LK method is proposed in Section \ref{s:CLK method}, where the PB data is augmented with an additional datum on the species' prevalence $\pi$. 
This makes the LI/LK methods comparable to the EM, SB and SC methods.

\section{A unified Constrained LK (CLK) method }
\label{s:CLK method}

From previous studies, we have found that the LI, LK, MAXENT and the conditional PPM share a similar likelihood function  i.e., Eq.~\ref{eqn:LK} and Eq.~\ref{eqn:approx_con}, which alone (without extra information) can only provide the relative probability (intensity) of presence. In order to obtain the absolute probability of presence, an extra information of the species' prevalence $\pi$ can be introduced as a constraint imposed on the optimization of this common likelihood function. In details, the CLK method maximizes the following (LK type of) likelihood function, 
\begin{eqnarray}
L_1(\beta)=\prod_{i=1}^{n_1} \frac{p(y=1|x_i,\beta)}{\frac{1}{n_0}\sum_{j=1}^{n_0} p(y=1|x_j,\beta)}, 
\label{CLK}
\end{eqnarray}
with the constraint, i.e., ${\frac{1}{n_0}\sum_{j=1}^{n_0} p(y=1|x_j,\beta)}=\pi_0$, where $\pi_0$ is the population prevalence that is assumed to be known in advance. 
Note that this is very different from just maximising the function of $\log L=\sum_{i=1}^{n_1} \log \frac{p_i}{\pi_0}$, since the constraint reduces the effective parameter space over which the maximization is performed. The statistical mechanism and efficiency underlying the CLK method is provided in Appendix C,  
where we have proved that the CLK is capable of estimating the true probability of presence, the same as the SB and SC methods. 

The LI, LK and the partial likelihood of the PPM (or MAXENT) would intrinsically have identification problems in solving their likelihood functions, if there is no prior knowledge of the species prevalence, and/or the structure of the function of the probability of presence.  In other words, these methods in general would generate multiple solutions of the absolute probability of presence, i.e., the relative probabilities of presence. 
By introducing the constraint, the CLK method forces the estimates from these methods to converge to the unique solution, which is just one of the multiple solutions obtained from the LI, LK and the MAXENT methods.
The CLK method provides a unification of 
the seemingly disparate methods discussed so far (SB, SC, EM, LI, LK, Lele, PPM and MAXENT). Each of these methods can be shown to be either equivalent to, or a special case of, the CLK method.

Firstly, LK, LI and Lele are special cases of the CLK method, when $\log p(y|x,\beta)$ is a nonlinear function and no constraint is used. 
If the RSPF conditions \cite{lele_weighted_2006,Solymos_Lele_2016} are not satisfied, using the logit-linear or other functions to fit without constraint fails to estimate the probability of presence \cite{phillips_estimating_2013}.
The inclusion of the additional information of $\pi$ in the CLK method fixes this problem, and enables the LI/LK methods to perform as well as SB, SC or EM method. 

Secondly, the CLK method has the same performance as the SB, SC and EM methods, when the logit link function is employed. However, unlike these methods, which were only derived for the logit function, the formulation of the CLK method is much simpler and can easily adapt to any type of link functions. 

Next, the PPM can be reviewed as a special case of the CLK method, when the log-linear function is used  for $p(y|x,\beta)$, and a constraint of $\frac{n_1}{|D|}$ is imposed on the denominator of Eq.~\ref{CLK}.
For a log-linear function function, i.e., $\log p(y|x,\beta)=\beta_0+\beta_1'x$, estimates of $\beta_1$ are the same for both the CLK method and the conditional PPM (equivalently the MAXENT), whereas the ratio of the two methods differ by a constant, i.e., the exponent of the difference between the two $\beta_0'$s \cite{fithian_finite-sample_2013}. 
It is the constraint that provides the estimate of the intercept in the log-linear model.  Similarly, MAXENT model is also a special case of the CLK method, using the logarithm function but without any constraint supplied.

Unlike all of these  previous methods, the CLK does not specify any particular link function; instead it can use any of the commonly used link functions, such as the logit, log or the complementary log-log functions. Interestingly, we will show in the Simulation section that using different link functions actually have little difference on estimating both the relative and the absolute probabilities of presence. 
The proposed CLK method is easy to implement, and users can choose any general-purpose nonlinear constraint optimization package in their preferred programming language. We have implemented the CLK method in R, and used the constraint optimization package `nloptr' \cite{NLOPTR}. The code of the new method is included in the Supporting Information.   

\section{Simulations}
 
In this section, the performance of the proposed CLK method is evaluated through  numerical simulations, using three commonly applied link functions, logit-linear, log-linear, and complementary log-log, denoted separately as CLK\_logit, CLK\_log and CLK\_clog. The CLK method can easily include other link functions. The large sample equivalence between the LI and LK methods is also demonstrated through these numerical experiments.
	
We consider eight species, with seven of them having the same probability functions of occurrence used in Phillips and Elith (2013) (see Table~\ref{table_species}). The extra species considered in our paper has the exponential distribution. The probability of presence $p(y=1|x)$ depends only on a single environmental covariate or explanatory variable $x$, and its value ranges uniformly between [0,1]. Five models were considered, i.e., LI, LK, CLK\_logit, CLK\_log and CLK\_clog. In our simulation,  no knowledge is assumed about the parametric structure of the true probability of presence, and we fit the data with the commonly used logit function for both the LI and LK methods. 
\begin{table}
 \caption{Probability of presence for eight simulated species}    
 \label{table_species}
 \begin{center}
  \begin{tabular}{c|c} \hline
   Simulated species &  Probability of presence \\ 
	& $P(y=1|x)$ \\ \hline
Constant & 0.3\\ 
Linear & $0.05+0.2x$ \\
Exponential &  $\exp(-4+4x)$\\
Quadratic & $0.5-1.333(x-0.5)^2$\\
Gaussian & $0.75\exp[-(4x-2)^2]$ \\
Semi-logistic & $8/(1+\exp[4-2x])$ \\
Logistic-1 & $1/(1+\exp[4-2x])$\\
Logistic-2 & $1/(1+\exp[4-8x])$ \\ \hline
\multicolumn{2}{l} \small{Note: $x$ is the single environmental covariate that is uniformly distributed on [0,1]} \\
\end{tabular}
 \end{center}   
\end{table}

\begin{figure}[htbp]
 \centerline{\includegraphics[width=4.5in]{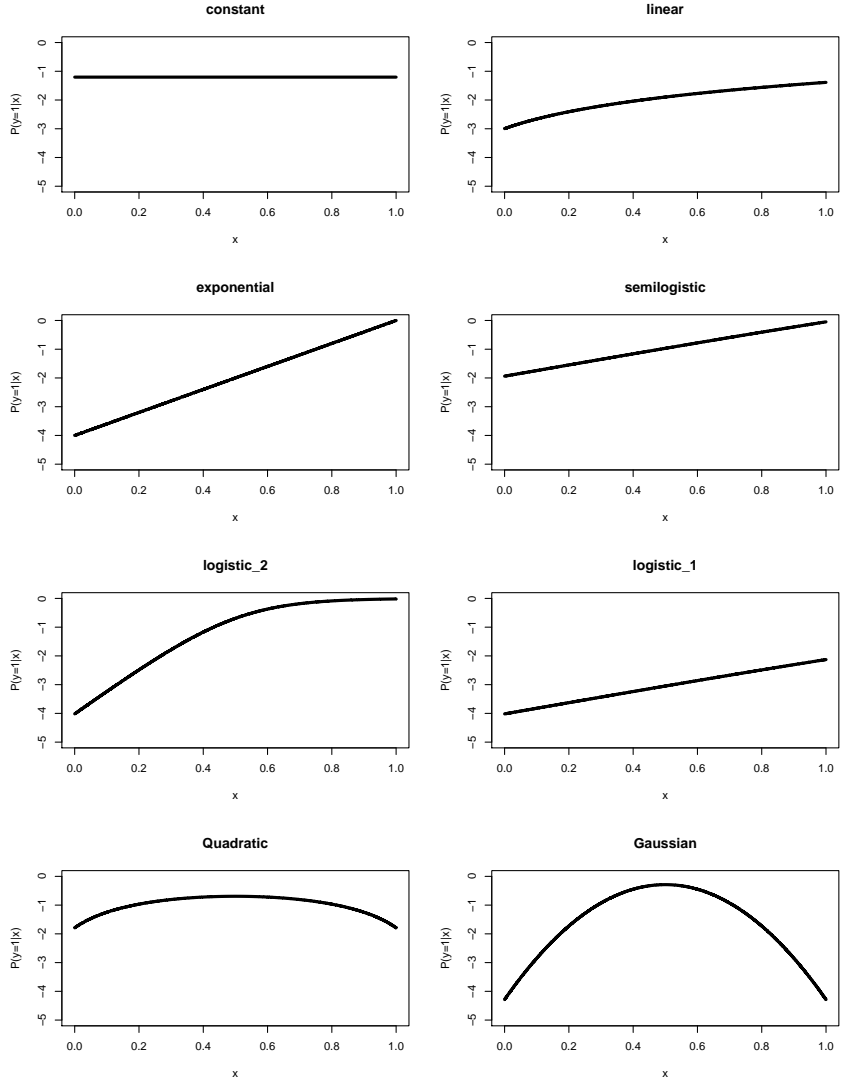}}
 \renewcommand{\baselinestretch}{1}
 \caption{The logarithm of each probability function in Table~\ref{table_species} is plotted, in order to verify the RSPF conditions provided by Lele and Keim (2006) and Solymos and Lele (2016), i.e, $\log p(y|x,\beta)$ being non-linear. Three graphs of the probabilities of presence are nonlinear and thus satisfy the RSPF conditions, and they are Logistic-2, Quadratic and Gaussian distributions.}
\label{figure_RSPF}
\end{figure}

We plot the logarithm of each probability function in Table~\ref{table_species} to verify the RSPF conditions listed in \citeA{lele_weighted_2006} and \citeA{Solymos_Lele_2016}. 
It is observed from Figure~\ref{figure_RSPF} that only Logistic-2, Quadratic and Gaussian distributions exhibit the required nonlinearity and appear  to satisfy the RSPF conditions, from the eight species.

For each species, 2,000 presence samples were drawn representing the locations of 2,000 observed individuals. Similarly 20,000 background samples were drawn. For each species, 100 simulations were run, and both the LI and LK methods were used to fit each simulation. The three CLK models were only fitted and plotted for one of the 100 simulations respectively, as all the 100 fits were very similar to each other for each CLK model. The fits were compared both visually (Fig.~\ref{figureI}) and using the root mean square (RMS) error (Fig.~\ref{figureII}) as the assessment statistics, against the true probability of presence. For the "Quadratic" and "Gaussian" species, quadratic terms of $x$ were added to fit the true probability. As the CLK method requires an estimate of the species' prevalence, we use the true prevalence as the estimate. 
Sensitivity analysis was also carried out by varying the true prevalence by $\pm 0.1$, and the results are reported in Fig.~\ref{figureI} as well.

We note that the numerical results of the LI and LK methods reported in Phillips and Elith (2013) appear different to those reported here, because parameters of these two methods are not identifiable in some of the simulations. In our simulations,  the identifiability was assessed by computing the reciprocal of the condition number, the ratio of the largest to the smallest eigenvalues of the Hessian  matrix. A ratio very close to zero (not exactly zero using the Hessian matrix as the estimate) indicates an identifiability issue. We arbitrarily chose 0.001 as the threshold to assess the identifiability for each simulation. The summary statistics (means and standard errors of the estimates) were computed for the adjusted intercept $\hat\beta_0$ and slope $\hat\beta_1$, after removing  unidentifiable simulations. In order to demonstrate the large sample equivalence between the LI and the LK methods, both methods were fitted with logit-linear and log-linear functions, and their summary statistics are shown in Table~\ref{table_LILKlogit} and Table~\ref{table_LILKlog} separately. Only the slope estimates are reported in Table~\ref{table_LILKlog}, because the intercept of the log-linear model is not identifiable for the LI and LK methods.

We also plotted the ratio of $\frac{p(y=1|x,\hat\beta)}{\hat \pi}$ for the three CLK methods, as well as the LK method fitted with log-linear and logit-linear functions respectively, shown in Fig.~\ref{figure_ratio}. 
The relative probabilities of the LK method fitted with the log and logit linear link functions were plotted for each of the 100 simulations, while the relative probabilities for the CLK methods were only plotted once due to the high similarities among the 100 replications.

The R code provided by Phillips and Elith (2013) facilitated our programming process. All model-fitting was carried out in R version 3.2.2 (R Core Team, 2016).

\section {Results} 

\begin{figure}[htbp]
 \centerline{\includegraphics[width=4.5in]{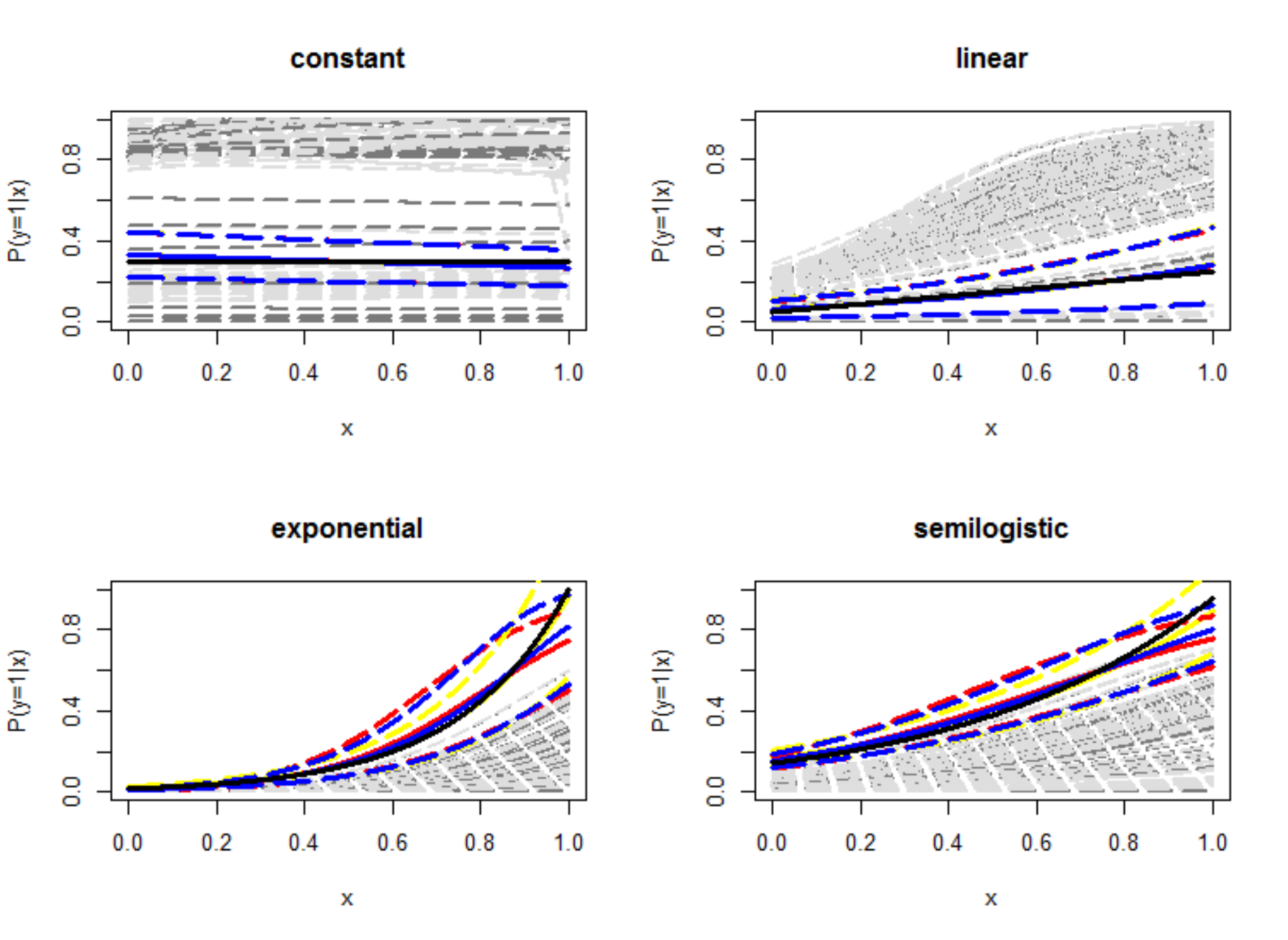}}
\centerline{\includegraphics[width=4.5in]{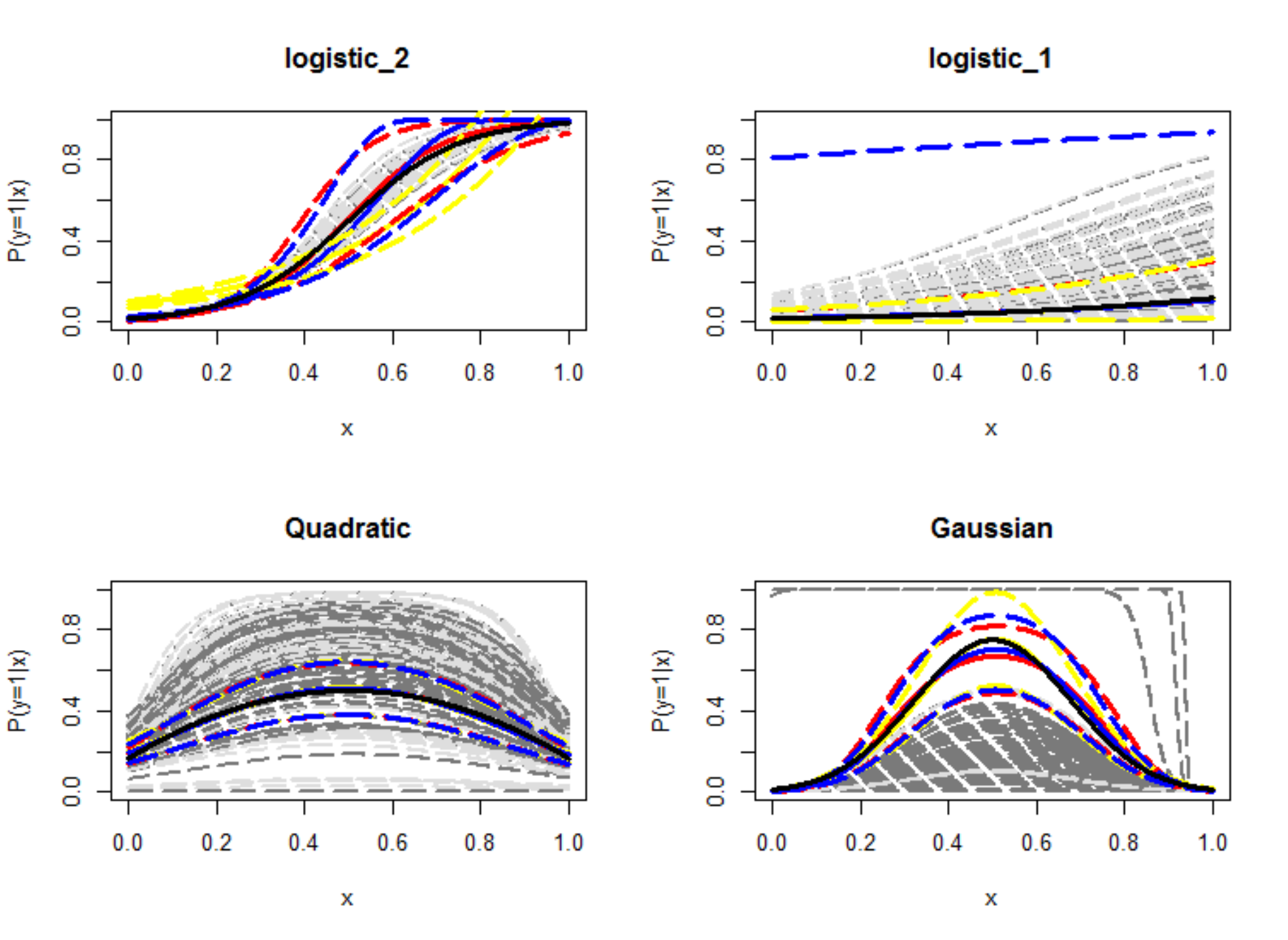}}
 \renewcommand{\baselinestretch}{1}
 \caption{LK and LI methods were fitted for each species with a replication of 100 times (two types of grey dotted lines), with the true probability given by the black line. CLK method was fit with the logit-linear function using the true prevalence (red lines), and logit function using the true prevalence $\pm 10\%$ (red dashed lines). CLK method was fit with the log-linear function using the true prevalence (yellow line), and log-linear function using the true prevalence $\pm 10\%$ (yellow dashed lines). CLK method was fit with the complementary loglog function using the true prevalence (blue lines), and using the prevalence $\pm 10\%$ (blue dashed lines).}
\label{figureI}
\end{figure}
%\floatplacement{figure_ratio}{H}

Firstly, we see in Fig.~\ref{figureI} that when the true species probability of presence is logit-linear in the case of Logistic-2, the LI/LK methods fit the data well, because the logit-linear function satisfies the RSPF conditions \cite{lele_weighted_2006, Solymos_Lele_2016}. 
In most other cases, both LI/LK methods have a wide spread for their estimates in the plots, which gives an indication of the non-identifiability of LI/LK methods in estimating the probability of presence. These happened because the probability functions for most simulated species do not satisfy the RSPF conditions (see Figure~\ref{figure_RSPF} for details). The Quadratic and Gaussian distributions do satisfy the RSPF conditions, however, the performance of the LI/LK model using the logit function were not good for fitting these two distributions. When the PB data is augmented with the species' prevalence, the CLK method closely approximates the true probability of presence, using the loglinear (red lines), logit (red lines) or the complementary loglog link functions (blue lines) (except for the species of Logistic\_1). 
\begin{figure}[htbp]
\centerline{\includegraphics[width=6.5in]{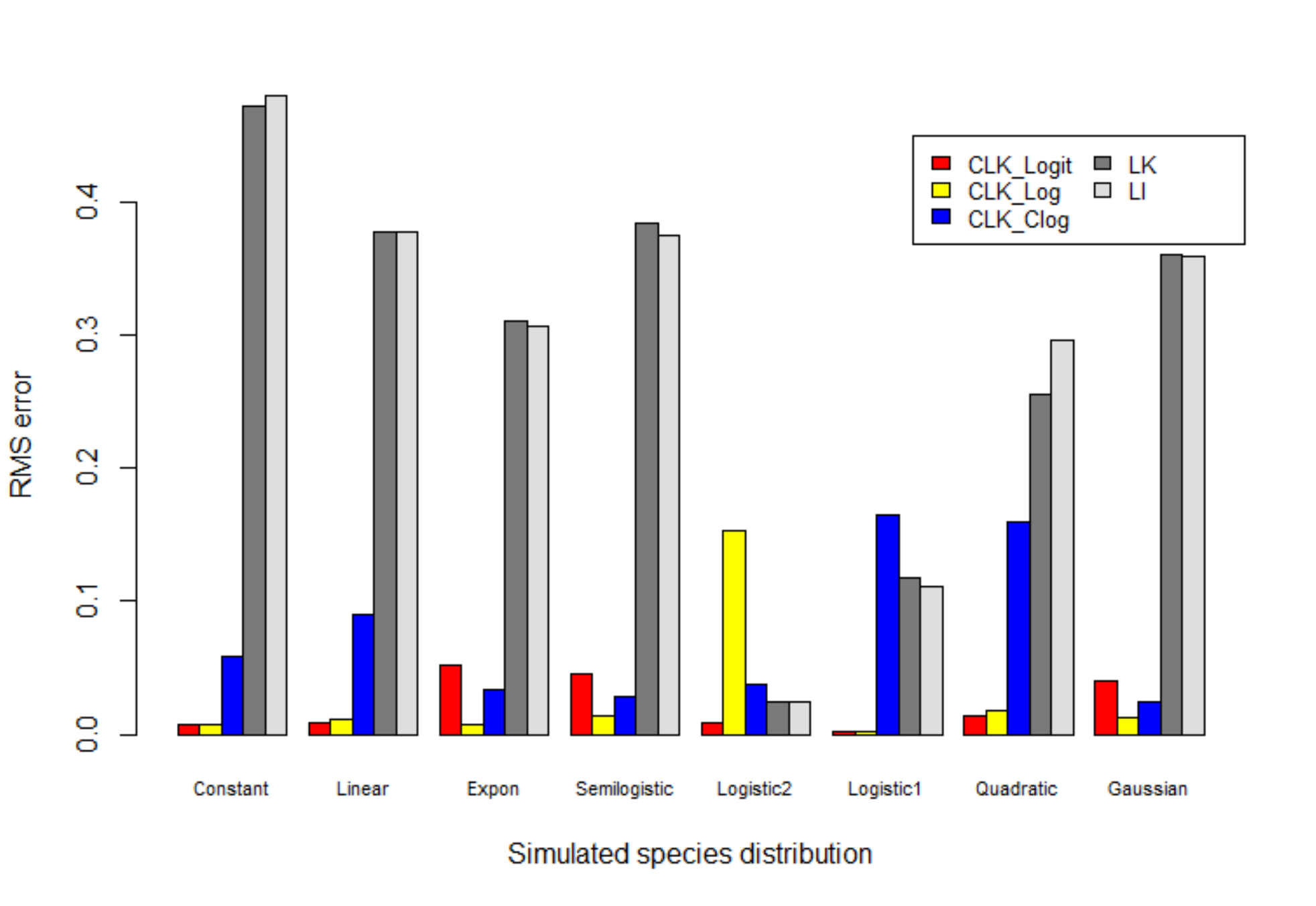}}
\caption{Root mean square (RMS) error of the LI, LK and the CLK method using the logit (CLK\_Logit), log-linear (CLK\_Log) and the complementary log-log (CLK\_Clog)functions.}
\label{figureII}
\end{figure}

Secondly we can see a close resemblance between the LI method (yellow lines) and LK method (purple lines) fitted with logit function respectively (Fig.~\ref{figureI}). This resemblance is further supported by the RMS errors of the two methods (see two grey column charts in Fig.~\ref{figureII}).  However, there still exists some discrepancy between the two methods for some simulated species, for example with the constant and quadratic distributions. 
The mean and standard errors of the estimates for the LI and LK methods are similar to each other in Table~\ref{table_LILKlogit}. The similar results of the LI and LK methods can also be seen  in Table~\ref{table_LILKlog}, where both methods were fitted with the log-linear functions. It is obvious that slope estimates between the LI and LK method are nearly the same. As the LK method is just the numerical approximation of the LI method, any difference between the two methods become less obvious as the number of background points increase.

\begin{table}
\centering
\caption{Mean of $\hat\beta_0$ and $\hat\beta_1$ for LI, LK and CLK, fitted with logit-linear function \hspace{\textwidth} (standard errors provided in parentheses)}
\label{table_LILKlogit}
\resizebox{\textwidth}{!}
{\begin{tabular} {l|cccccc} 
\hline
 & LK-{$\hat\beta0$} & LI-${\hat\beta0}$ & LK-${\hat\beta1}$ & LI-${\hat\beta1}$ &CLK-${\hat\beta0}$ & CLK-${\hat\beta1}$ \\ \hline 
Constant & 3.599 (3.233) & 3.508 (3.242) & 3.758 (9.200) & 1.053 (7.233) &-0.858 (0.078) & 0.0211 (0.0157) \\ 
Linear & -1.529 (0.271) & -1.527 (0.268) & 3.119 (0.767) & 3.128 (0.770) & -2.628 (0.085) & 1.634 (0.147)\\ 
Exponential & -5.716 (0.559) & -5.724 (0.574) & 4.356 (0.289) & 4.355 (0.290) & -4.550 (0.212) & 5.664 (0.316) \\ 
Semilogit & -2.958 (0.550) & -2.963 (0.558) & 2.422 (0.305) & 2.421 (0.307) & -2.048 (0.113) & 3.452 (0.218) \\
Logistic1 & -2.821 (0.539) & -2.822 (0.541)  & 2.491 (0.467)  & 2.491 (0.467) & -3.991 (0.084) & 1.985 (0.132) \\
Logistic2 & -4.055 (0.243) & -4.056 (0.243) & 8.073 (0.777) & 8.074 (0.775) & -4.050  (0.223)& 8.105 (0.462) \\
Quadratic & -0.275 (0.473) & -0.280 (0.474) & 6.753 (8.228) & 6.963 (8.377) & -1.489 (0.136)& 6.036 (0.669) \\
Gaussian & -2.909 (0.615)  & 2.892 (0.528) & 4.178 (6.343) & 4.232 (6.324) & -5.038 (0.298)& 23.001 (1.319)\\ \hline
\end{tabular}}
\end{table}

\begin{table}
\caption{Mean of $\hat\beta_1$ for LI, LK and the CLK, fitted with log-linear function \hspace{\textwidth} (standard errors provided in parentheses)}
\label{table_LILKlog}
\begin{center}
\begin{tabular}{c|ccc}  \hline
 & LK1-${\hat\beta1}$ & LI1-${\hat\beta1}$ & CLK\_log-${\hat\beta_1}$\\ \hline 
Constant &  0.015 (0.109) & 0.015 (0.109))& 0.015 (0.109)) \\ 
Linear & 1.367 (0.121) & 1.370 (0.121)& 1.367 (0.121)  \\ 
Exponential & 3.992 (0.183)& 3.995(0.184)  & 3.992 (0.183) \\ 
Semilogit &  1.909 (0.113) & 1.910 (0.113) & 1.909 (0.113)  \\
Logistic1 & 1.869 (0.124)& 1.871 (0.124)& 1.869 (0.124)   \\
Logistic2 & 2.767 (0.109) & 2.802 (0.109) & 2.766 (0.109) \\
Quadratic & 3.813 (0.440)& 3.819(0.440) & 3.813 (0.440)  \\
Gaussian & 16.046 (0.825) &  16.047 (0.824)& 16.046 (0.825)\\ \hline
\end{tabular}
\end{center}
\end{table}

Upon examining Table~\ref{table_LILKlogit}, it is hard to see a clear and simple relationship between the LI/LK and the CLK estimates obtained when fitted with the logit-linear function. The estimates from both the LI and LK methods in general have higher standard errors compared to the CLK estimates. For some species such as the Quadratic or the Gaussian distribution, both the intercept and the slope have significantly large standard errors that would lead to possible rejection of the influential covariate, if we were to use the LI and LK methods to make statistical inference. However when all methods were fitted with the log-linear functions in Table~\ref{table_LILKlog}, not only are the slope estimates of the LI and LK methods nearly the same, but they are also the same for the CLK method. The resulting relative probabilities of presence from these three models are all proportional to the true probability of presence, by a ratio of $1/\log \hat\beta_0$, estimated from the CLK method.  Meanwhile in most of our simulated species, the estimates fitted by a log-linear function in general have a smaller variation compared to the estimates fitted with either a logit or complementary loglog functions. The performance of the complementary loglog functions is in particular poor,  when the true probability is very low as in the species of Logistic\_1 (see Fig.~\ref{figureI} and \ref{figureII}).

\begin{figure}[htbp]
\centerline{\includegraphics[width=4.5in]{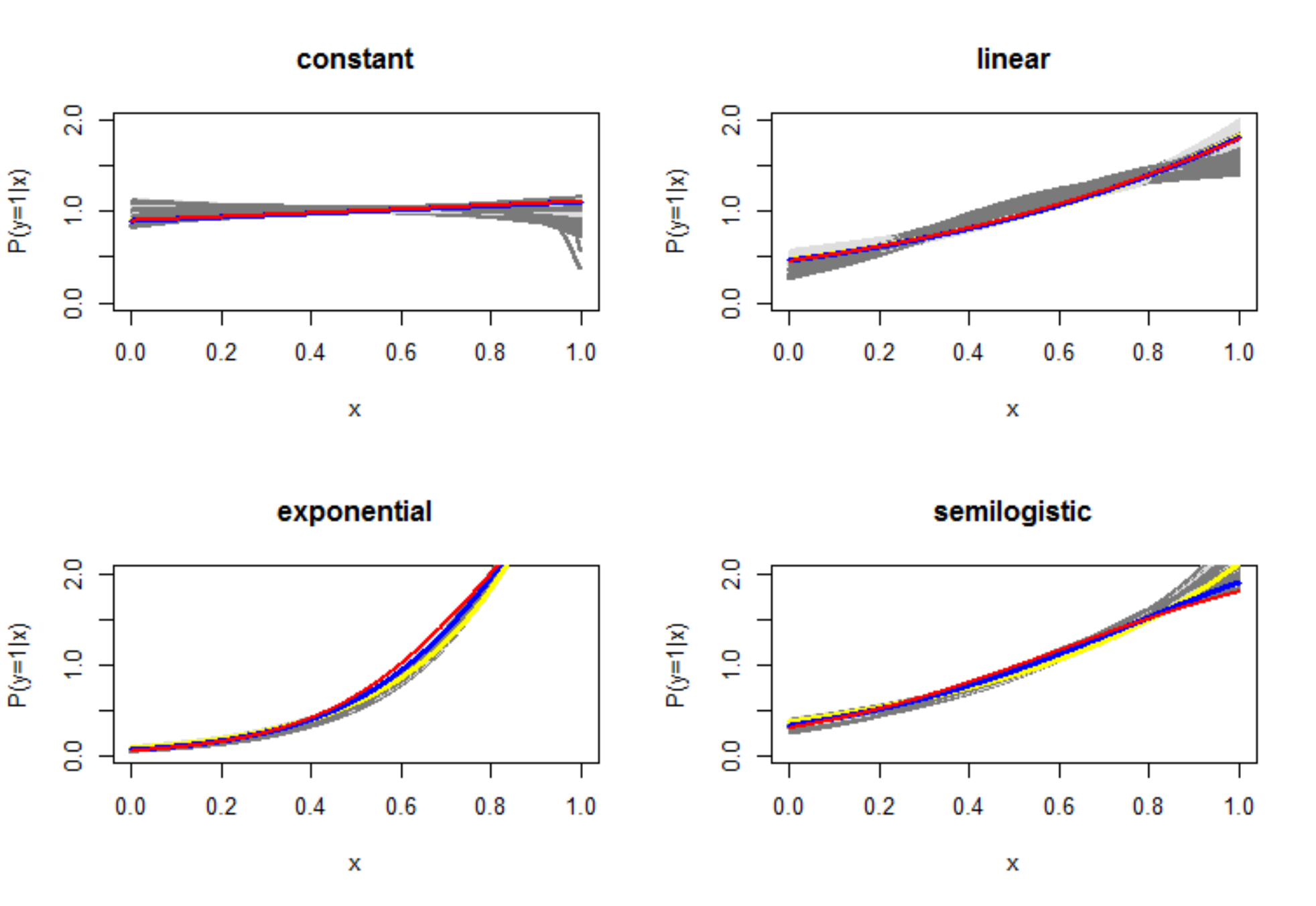}}
\centerline{\includegraphics[width=4.5in]{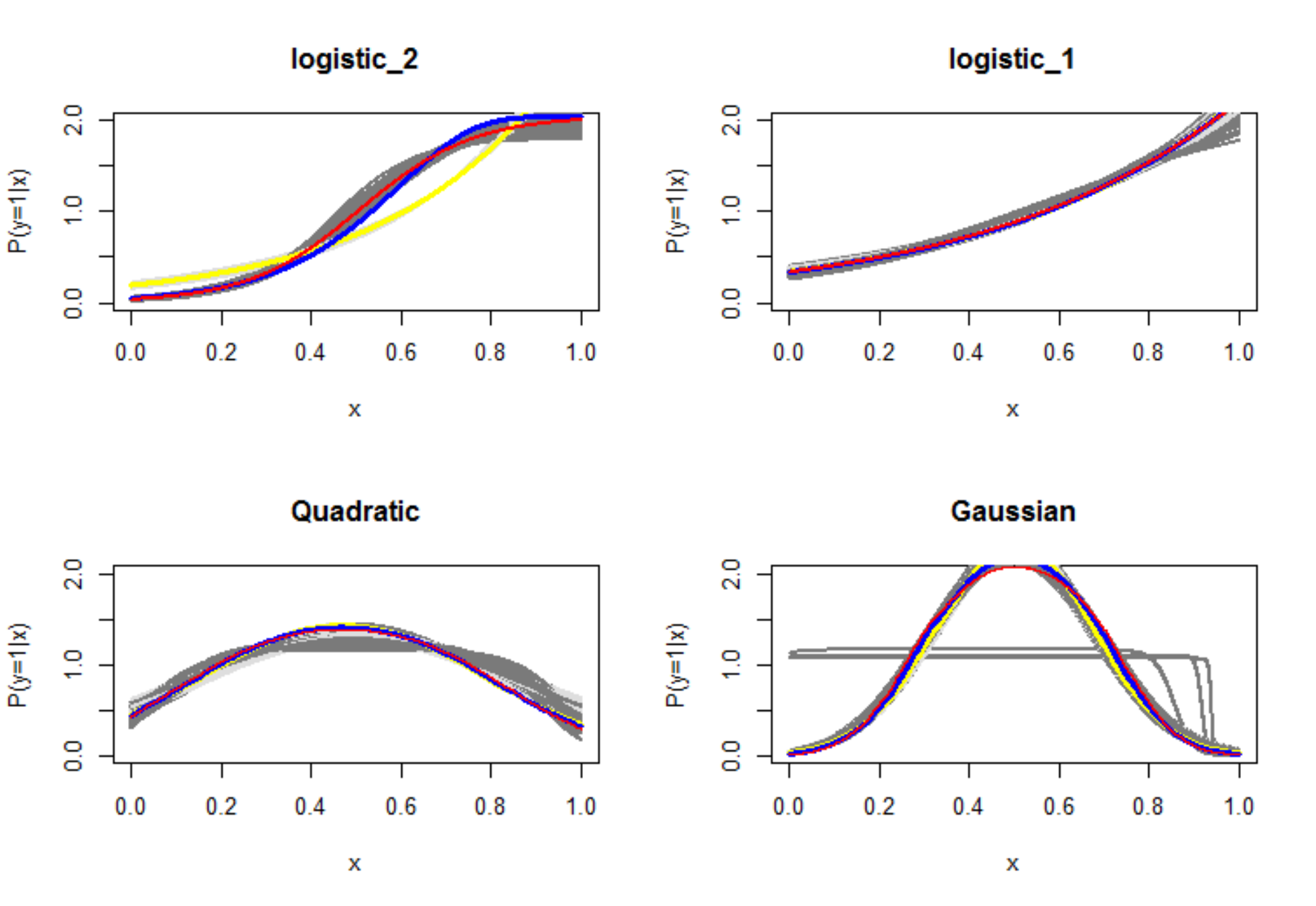}}
\caption{The ratio between the estimated probability of presence $p(y=1|x,\hat\beta)$ and the estimated population prevalence $\hat \pi$, are fitted with the LK method using logit (grey 1 line) and log-linear function (grey 2 line) over 100 simulations. The ratio is also fitted with the CLK\_Logit (red line), CLK\_Log (yellow line) and CLK\_Clog methods (blue line), using one randomly selected simulation (due to resemblance among replications).}
\label{figure_ratio}
\end{figure}

Although it is hard to see what the LK or LI method have estimated in Table~\ref{table_LILKlogit}, this ambiguity, however, becomes clear when we plot the relative probability of presence, i.e., the ratios $\frac{p(y=1|x,\hat\beta)}{\hat \pi}$ of the LK estimates fitted with both the logit and log-linear functions (Fig.~\ref{figure_ratio}). Comparing these ratios with the CLK estimates, we see that these ratios are all similar to each other, regardless of the functional form of the link function and which type of likelihood (full vs the conditional) have been used to fit the PB data. It further confirms that the LK/LI method can provide a good estimate of the relative probability of presence, when no extra information is available on either the RSPF conditions \cite{lele_weighted_2006, Solymos_Lele_2016} or the species' prevalence. Also there are some `erratic' curves observed for the LI and LK estimates in Fig.~\ref{figureI} and ~\ref{figure_ratio} for the species with a Gaussian distribution. These estimates were again simply caused by the non-identifiability problem in the LI and LK methods.

%\newpage
%\floatplacement{figure_ratio}{H}
We have also noticed an overall better fit in our simulations using the log-linear function compared to the logit and the complementary loglog functions. This may be due to the fact that most probability functions in our simulation belong to the exponential family. Meanwhile in our numerical studies, we have also observed that when the predetermined prevalence rate $\pi$ varies from the true prevalence $\pi_0$, the estimated probabilities of presence at each location from the CLK method using log-linear function still provides a consistent ranking, for the true probabilities of presence at each location. However, when the supplied estimated prevalence rate is too high, this is not true for the logit and complementary loglog functions. The estimated probabilities would level off at a probability close to 1, and therefore not able to give a correct ranking. This may be another reason that the log function is used as the link function in the popular MAXENT method.   

\section{Discussion}

In this paper we have revisited some commonly used methods for modelling species probability of presence with PB data. These methods include the LI \cite{lancaster_case-control_1996}, LK \cite{lele_weighted_2006,royle_likelihood_2012}, Lele \cite{lele_new_2009}, MAXENT \cite{phillips_modeling_2008}, point process models \cite{warton_poisson_2010,chakraborty_point_2011}, EM \shortcite{ward_presence-only_2009}, SB \cite{Phillips_SB_2011} and SC \cite{steinberg_estimating_1992} methods. It is not fair to compare their performance, because these methods are formulated on different conditions and prior information. When there is no information available on the conditions, we can conclude that all these methods for estimating the relative probability of presence, regardless of being well known or not, are essentially the same. Furthermore, these methods also have similar performance in estimating the absolute probability of presence, when the same additional information is provided. %depending  }well known and routinely applied methods for analyzing PB data, including logistic regression, are closely linked to each other. 
Firstly, we have shown that it is the conditional/partial likelihood that is actually employed for modelling PB data, which alone (without the extra information) can only be used to make inference about the relative probability of presence. The LI, LK, Lele, MAXENT and the conditional PPM model, were built upon the conditional/partial likelihood. 
 Other methods, such as the Poisson generalized linear regression, logistic regression and the PPM, can only estimate the probability of reporting (as opposed to the probability of presence), due to the lack of appropriate information on the true population prevalence or number of true presences \cite{fithian_finite-sample_2013,dorazio_accounting_2014,renner_point_2015}. 
 
Another important contribution of the paper is to propose a constraint CLK method. It makes the LI/LK approaches comparable to other well performing methods (SC, EM and SB), when the parametric RSPF conditions are not satisfied. 
Under this circumstance, the LI, LK and the conditional likelihood of the PPM (or MAXENT) methods intrinsically have identification problem in estimating all the parameters relevant to the true probability of presence (or intensity). 
The introduction of the constraint in the CLK method guarantees a unique estimate for the probability of presence, which equals the true population prevalence if the supplied prevalence is equal to the true one.  This unique estimate is just one of the multiple solutions obtained from the LI, LK and the MAXENT methods.

One may argue that the CLK method requires the population prevalence $\pi$ which is sometimes hard to obtain or estimate in practice. For our purposes, the CLK method proposed in this paper serves more as a technical generalisation tool to gain insight into modelling presence-background data, and to look at the connection of seemingly different methods. On the other hand, the information of population prevalence can be obtained from either pilot studies or other types of data, for example, the presence-absence (PA) survey data or the complementary expert map. There have been a few recent studies on the combination of PB and PA data \cite{dorazio_accounting_2014,fithian_bias_2015,Koshkina_2017}. These combined methods can estimate the absolute probability of presence successfully, by gaining the information of population prevalence from the PA
data.

It was shown in the previous section that the PPM cannot estimate the absolute intensity of presence, given $n_1$ is not a true reflection of the number of presences in study area. In order to estimate the true intensity for the PPM, it can simply maximize the conditional PPM likelihood Eqn~\ref{eqn:con}, with the constraint of $\Lambda(D)= \mbox{true number of presences over the study area}$, the same idea behind the CLK method. 

%\backmatter
\section*{Acknowledgments}
The support of the Australian Research Council grant DP150102472 is gratefully acknowledged.  
\vspace*{-8pt}

%\section*{Supplementary Materials}

%\section*{Supporting Information}

%Details of supporting information are provided below

%\noindent\textbf{Appendix A}: Equivalence of EM, SB and the LI methods

%\noindent\textbf{Appendix B}: Equivalence between the Lele (2009) and LI (1999) methods

%\noindent\textbf{Appendix C}: Statistical mechanism underlying the constrained LK(CLK) method

%\noindent R code for the CLK method and the simulations. 

\renewcommand{\APACrefYearMonthDay}[3]{\APACrefYear{#1}}
\bibliography{library}

\appendix
\newpage
\section*{Appendix A: Equivalence of EM, SB and the LI methods}
\setcounter{equation}{0}
\renewcommand\theequation{A.\arabic{equation}}

In the following, we demonstrate how the Expectation-Maximisation (EM) method \shortcite{ward_presence-only_2009} and the scaled binomial loss model (SB) \cite{Phillips_SB_2011} are essentially the same as the LI method through simple mathematical derivations.

\citeA{ward_presence-only_2009} let $z = 1$ and $z =
0$ denote the observed presences and background data, respectively. Note that when $z = 1$, we know $y = 1$. However, when $z = 0$ we do not know whether $y = 0$ or $y = 1$. The EM method proposed the following likelihood for the presence-background data \cite{ward_presence-only_2009}:
\begin{eqnarray}
 L(\eta|z,X) & = &\prod_i P(z_i|s_i=1,x_i) \hspace{0.5cm} ({\mbox{\boldmath{$EM$}}}) \\ \nonumber
  &=& \prod_i \left(\frac{\frac{n_1}{\pi n_0} e^{\eta(x_i)}} {1+(1+\frac{n_1}{\pi n_0})e^{\eta(x_i)}}\right)^{z_i} \left(\frac{1+e^{\eta(x_i)}} {1+(1+\frac{n_1}{\pi n_0})e^{\eta(x_i)}}\right)^{1-z_i}.
	\label{EM} 
\end{eqnarray}
Here $n_0$ is the number of observed presences denoted by $z=1$, and $n_1$ the number of background points denoted by $z=0$. The
notation $s = 1$ is a construct of case–control modeling and indicates that this observation is in the presence-background data
sample. The logit link function is used to model the true probability of presence, i.e., $\log\frac{p(y=1|x)}{1-p(y=1|x)}=\eta(x)$. 
As the information on the true presence $y$ is missing, direct maximisation of this likelihood is difficult. The expectation-maximization (EM) technique is implemented on the full likelihood of both the true and observed presences, with the missing $y$ imputed with its expectation. 

Now examine the SB method, in which the probability used in the likelihood function is defined as $P_{UA}(s=1|x)=\frac{1}{1+r+\exp(-\eta(x)+\ln r)}$ \cite{Phillips_SB_2011}, where $r$ equals $\frac{1-f_p}{f_p}\pi$ through the sampling probability of the presence points $f_p$. The sampling probability $f_p$ in the SB method can be rewritten as $f_p=\frac{n_0}{n_1+n_0}$, and it therefore gives $r=\frac{n_0}{n_1}\pi$. The probability $P_{UA}(s=1|x)$ can be re-written as 
\begin{eqnarray}
P_{UA}(s=1|x) \hspace{0.2cm} ({\mbox{\boldmath{$SB$}}}) &= &\frac{1}{1+r+\exp(-\eta(x)+\ln r)} \nonumber \\ &= & \frac{1}{1+\frac{\pi n_0}{n_1}+\frac{\pi n_0}{n_1}e^{-\eta(x)}} \nonumber \\ & = &\frac{\frac{n_1}{\pi n_0} e^{\eta(x)}} {1+(1+\frac{n_1}{\pi n_0})e^{\eta(x)}}.
\label{SB}
\end{eqnarray}
It is obvious that $P_{UA}(s=1|x)$ used in the SB method is exactly the same as $P(z|s=1,x)$ of the EM method. Instead of working indirectly on the likelihood of the observed data (as the EM method), the SB method  directly maximizes the observed likelihood function from the outset, by using a modification of the standard binomial loss function.  

 In the LI method, each observed presence is drawn uniformly with the probability $h$, the same as $f_p$ defined in the SB method. The likelihood function $L(\beta,\pi,h)$ is constructed on the probability $R_{1n}$ through $L(\beta,\pi,h)=\prod_i R_{1n}(\beta,\pi,h)^{z_i}(1-R_{1n}(\beta,\pi,h))^{1-z_i}$, where $R_{1n}=\frac{(h/\pi) P(y=1|x,\beta)}{(h/\pi) P(y=1|x,\beta)+1-h}$ \cite{lancaster_case-control_1996}. When $p(y=1|x,\beta)$ takes the same logit-linear function as the EM and SB methods, i.e., $P(y=1|x,\beta)=\frac{e^{\eta(x)}}{1+e^{\eta(x)}}$, one finds that 
\begin{equation}
 R_{1n} \hspace{0.2cm}({\mbox{\boldmath{LI}}})=\frac{\frac{n_1/\pi}{n_1+n_0} \frac{e^{\eta(x)}}{1+e^\eta(x)}} {\frac{n_1/\pi}{n_0+n_1}\frac{e^\eta(x)}{1+e^\eta(x)}+\frac{n_0}{n_0+n_1}}=\frac{\frac{n_1}{\pi n_0} e^\eta(x)} {1+(1+\frac{n_1}{\pi n_0})e^\eta(x)}.
\end{equation}

Obviously $R_{1n}(\mbox{\boldmath{$LI$}})=P(z|s=1,x)(\mbox{\boldmath{$EM$}})=P_{UA}(s=1|x) (\mbox{\boldmath{$SB$}})$, i.e. the probabilities on which the likelihood functions were formulated, are the same for these three seemingly different methods. The difference lies in the extra information required: the SB and EM methods need a pre-determined value of $\pi$, while the LI method treats $\pi$ as one of the unknown parameters. In order for the LI method to identify $\pi$, the identifiability conditions listed in \citeA{lele_weighted_2006} and \citeA{Solymos_Lele_2016} have to be satisfied. The numerical examples in \citeA{lancaster_case-control_1996} paper work well, because they satisfy these parametric identifiability conditions. 

\section*{Appendix B: Equivalence between the Lele (2009) and LI (1999) methods}
\setcounter{equation}{0}
\renewcommand\theequation{B.\arabic{equation}}

Lele's partial likelihood is given in Lele (2009; Eq.~2) as:

\begin{equation}
PL(\beta)=\prod_{i=1}^N \frac{w\pi(X_i^U,\beta)}{w\pi(X_i^U,\beta)+(1-w)P(\beta)} \prod_{j=1}^M\frac{(1-w)P(\beta)}{w\pi(X_j^A,\beta)+(1-w)P(\beta)} 
\label{PL}
\end{equation}

Before we show the equivalence between this partial likelihood (PL) in \citeA{lele_new_2009} and that of \citeA{lancaster_case-control_1996}, we summarized the comparable notations used by these differ approaches  

\begin{table}
\caption{The table summarizes the key symbols used in Lele (2009) and our paper, making them comparable to each other.}
\begin{center}
\label{table_appendixb}
\begin{tabular} {c|c|c}
 \textbf{Definition} & \textbf {Lele (2009)} & \textbf{Our paper}  \\ \hline 
Probability of presence &$p(y_i=1|x_i,\beta)$ & $\pi(X_i,\beta)$ \\ \hline
Population prevalence & $P(\beta)$ & $\pi$ \\ \hline 
Number of presences & $N$ & $n_1$ \\ \hline
Number of background & $M$ & $n_0$ \\\hline
Sampling probability & $w=\frac{N}{N+M}$ & $h=\frac{n_1}{n_1+n_0}$ \\ \hline
\end{tabular}
\end{center}
\end{table} 

The likelihood function $L_1(\beta,\pi,h)$ in Lancaster and Imbens (1999) is constructed from the probability $R_{1n}$ through $L_1(\beta,\pi,h)=\prod_i R_{1n}(\beta,\pi,h)^{z_i}(1-R_{1n}(\beta,\pi,h))^{1-z_i}$, where $R_{1n}=\frac{(h/\pi) P(y=1|x,\beta)}{(h/\pi) P(y=1|x,\beta)+1-h}$. Using the comparable notations in Table~\ref{table_appendixb}, we can easily rewrite the partial likelihood function of Lele (2009) as 
\begin{eqnarray}
PL(\beta)&=&\prod_{i=1}^N \frac{w\pi(X_i^U,\beta)}{w\pi(X_i^U,\beta)+(1-w)P(\beta)} \prod_{j=1}^M\frac{(1-w)P(\beta)}{w\pi(X_j^A,\beta)+(1-w)P(\beta)} \nonumber \\
&=& \prod_{i=1}^{n_1} \frac{hp(y=1|x_i,\beta)}{hp(y=1|x_i,\beta)+(1-h)\pi} \prod_{j=1}^{n_0} \frac{(1-h)\pi}{hp(y=1|x_j,\beta)+(1-h)\pi} \nonumber\\ 
&=& \prod_{i=1}^{n_1} \frac{(h/\pi)p(y=1|x_i,\beta)}{(h/\pi)p(y=1|x_i,\beta)+(1-h)} \prod_{j=1}^{n_0} \frac{(1-h)}{(h/\pi)p(y=1|x_j,\beta)+(1-h)} \nonumber \\
&=&\prod_i R_{1n}(\beta,\pi,h)^{z_i}(1-R_{1n}(\beta,\pi,h))^{1-z_i} \nonumber \\
&=&L_1(\beta,\pi,h). 
\end{eqnarray}
Therefore both the Lele (2009) and the LI methods are based on  exactly the same likelihood functions.

\section*{Appendix C: Statistical mechanism underlying the constrained LK(CLK) method}
\setcounter{equation}{0}
\renewcommand\theequation{C.\arabic{equation}}

If we look further using Lagrange multipliers for the proposed CLK method, the Lagrange function is 
\begin{equation}
L(p_i,\lambda)=\sum_{i=1}^{n_1}\{\log p_i-\log \pi_0\}-\lambda(\frac{\sum_{i=1}^{n_0} p_i}{n_0}-\pi_0),
\label{Lagrange}
\end{equation} 
with the constant Lagrange multiplier $\lambda$. Calculate the gradient of (\ref{Lagrange}) with respect to $p_i$ and $\lambda$ respectively,
\begin{eqnarray*}
\nabla_{p_i,\lambda} L(p_i,\lambda) &=& (\frac{\partial L}{\partial p_i}, \frac{\partial L}{\partial \lambda}) \\ &=&  (\frac{n_1}{p_i}-\frac{\lambda}{n_0},\frac{\sum_{i=1}^{n_0} p_i}{n_0}-\pi_0).
\end{eqnarray*}
Solving $\nabla_{p_i,\lambda} L(p_i,\lambda)=0$, shows that the estimate of the probability of presence is equal to the population prevalence, i.e., $\hat{p_i}=\pi_0$.

\subsection*{Likelihood estimates of the SB and the SC methods}
The loglikelhiood of the SB method \cite{Phillips_SB_2011} is  
\begin{eqnarray*}
U(p_i)=\sum_{i=1}^{n_1} \log P_{UA}+\sum_{i=1}^{n_0} \log(1-P_{UA}).
\end{eqnarray*}
Here $p_i$ is  the logit function $p_i=\frac{1}{1+\exp(-\eta(x))}$, and $P_{UA}=\frac{1}{1+r+\exp(-\eta(x)+\ln r)}=\frac{1}{1+r/p_i}$. Taking the derivative of $U(p_i)$ with respect to $p_i$, setting it to zero, and the score function for $p_i$ becomes $n_0 P_{UA}=n_1(1-P_{UA})$, i.e., $\frac{n_0}{1+r/p_i}=\frac{n_1r/p_i}{1+r/p_i}$. In \citeA{Phillips_SB_2011}, $r$ is defined as $r=\frac{1-f_p}{f_p}\pi_0$, which is equivalent to $\frac{n_0}{n_1}\pi_0$, given the sampling proportion of the presence only points $f_p=\frac{n_1}{n_1+n_0}$. It therefore yields $\hat{p_i}=\frac{n_0}{n_1}r=\pi_0$.

As for the SC method \cite{steinberg_estimating_1992}, the log-likelihood function is 
\begin{eqnarray*}
L(p_i)=\frac{1}{n_0}\sum_{i=1}^{n_0}\log(1-p_i)+\frac{\pi_0}{n_1}\sum_{i=1}^{n_1}\log\frac{p_i}{1-p_i}.
\end{eqnarray*}
Similarly, solving the score function for $p_i$ leads to the estimate of $\hat{p_i}=\pi_0$. 

Therefore the proposed CLK method obtains the same estimate as the SC \cite{steinberg_estimating_1992} and SB \cite{Phillips_SB_2011} methods for $p_i$, which are all estimated to be equal to the predetermined probability of prevalence $\pi_0$. 
 
\label{lastpage}
\end{document}